\documentclass[11pt]{article}
\usepackage{amsfonts,amssymb,amsmath}
\usepackage{stmaryrd}

\def \TITLE {Talk on Anomalies in QFT and Cohomologies}

\usepackage{ulem}

\newcommand{\beq}{\begin{equation}}
\newcommand{\eeq}{\end{equation}}
\newcommand{\beqa}{\begin{eqnarray}}
\newcommand{\eeqa}{\end{eqnarray}}
\newcommand{\nn}{\nonumber \\}
\def \podr {&& \hspace{-15pt}}

\def \x {{\mathrm{x}}}
\def \xx {\mbf{\mathrm{x}}}

\def \y {{\mathrm{y}}}

\def \rr {\mbf{\mathrm{r}}}

\def \rR {\mathrm{r}}

\def \sS {\mathrm{s}}

\def \Ss {\mathrsfs{S}}
\def \Dd {\mathrsfs{D}}
\def \R {{\mathbb R}}
\def \C {{\mathbb C}}

\def \N {{\mathbb N}}

\def \id {\text{\rm id}}
\def \di {\partial}
\def \mz {\bigr\backslash\hspace{1pt}\{\Mbf{0}\}}

\def \RDF {Q}

\DeclareMathAlphabet{\mathbbm}{U}{bbm}{m}{n}

\DeclareSymbolFont{ltrs}     {OT1}{pzc}{m}{it}
\DeclareSymbolFont{ltrsa}     {OMS}{cmsy}{m}{n}
\DeclareSymbolFont{ltrsA}{U}{txmia}{m}{it}
\DeclareSymbolFont{symbolsC}{U}{txsyc}{m}{n}
\DeclareSymbolFont{ltrsB}{U}{rsfs}{m}{n}

\DeclareSymbolFontAlphabet{\mfrak}{ltrsA}
\DeclareMathAlphabet{\mathpzc}{OT1}{pzc}{m}{it}
\DeclareMathAlphabet{\mathrsfs}{U}{rsfs}{m}{n}

\def \La {
\left\langle \!\!{\,}^{\mathop{}\limits_{}}_{\mathop{}\limits^{}}\right.}
\def \Ra {
\left. \!\!{\,}^{\mathop{}\limits_{}}_{\mathop{}\limits^{}}\right\rangle}
\newcommand{\vrestr}[2]{\!\left.\raisebox{#1}{$\,$}\!\right|_{\,
\raisebox{1pt}{\small \(#2\)}}}

\newcommand{\mbf}[1]{\ensuremath{\mathchoice
                    {\mbox{\boldmath$\displaystyle\mathbf{\mathit{#1}}$}}
                    {\mbox{\boldmath$\textstyle\mathbf{\mathit{#1}}$}}
                    {\mbox{\boldmath$\scriptstyle\mathbf{\mathit{#1}}$}}
                    {\mbox{\boldmath$\scriptscriptstyle\mathbf{\mathit{#1}}$}}}}

\newcommand{\Mbf}[1]{\ensuremath{\mathchoice
                    {\mbox{\boldmath$\displaystyle\mathbf{#1}$}}
                    {\mbox{\boldmath$\textstyle\mathbf{#1}$}}
                    {\mbox{\boldmath$\scriptstyle\mathbf{#1}$}}
                    {\mbox{\boldmath$\scriptscriptstyle\mathbf{#1}$}}}}

\newcounter{Theorem}
\newcounter{Definition}
\newcounter{Remark}

\newcounter{tmpc}
\newlength{\tmplenght}
\newlength{\tmplenghta}
\newlength{\tmplenghtb}
\newlength{\tmplenghtc}

\def\DP{\Dd'}
\def\CI{\mathcal{C}^{\infty}}

\def\du{\hspace{2pt}\text{\tiny $\vee$}}

\newcommand{\OM}[1]{\Omega^{#1}}

\newcommand{\HOM}[1]{H^{#1}}

\def\Scdeg{\text{\rm Sc.{\hspace{1pt}}d.}}

\def \Lccl {\omega}

\def \Sccl {\gamma}
\def \sccl {\Gamma}
\def \bsccl {\Mbf{\Gamma}}

\def\Q{\mathbb{Q}}

\def\HF{\Mbf{F}}
\def\HES{\Mbf{E}}

\def\PA{{\mathrsfs{O}\hspace{-1pt}}}

\def\RMA{R}
\def\PRT{\mathfrak{P}}

\def\SRMA{{\mathop{\RMA}\limits^{\text{\tiny $\bullet$}}}}

\def\PRMA{P}

\def\NORM{\text{\it n.\hspace{1pt}f.}\hspace{1pt}}

\def\Rdf{Q}

\usepackage{ulem}

\def\SLccl{{\mathop{\Lccl}\limits^{\text{\tiny $\bullet$}}}}

\newcommand{\EMPHTH}[1]{{\rm #1}}

\def\spwedge{\mathop{\wedge}\limits^{\circ}}

\def\PAE{\hspace{2pt}\widetilde{\hspace{-2pt}\PA\hspace{2pt}}\hspace{-2pt}}

\title{Talk on Anomalies in Quantum Field Theory and Cohomologies
	of Configuration Spaces\thanks{%
	Talks given at the Conference on Algebraic and Combinatorial Structures
	in Quantum Field Theory, Carg\`ese, March 31, 2009
	and at the Conference
  ``Algebraic Methods in Quantum Field Theory''
  50 Years of Mathematical Physics in Bulgaria
  on the occasion of the 75th anniversary of Ivan Todorov,
	Sofia, May 15, 2009.
  Based on the paper: arXiv:0903.0187}}

\author{Nikolay M. Nikolov}

\begin{document}

\maketitle

\thispagestyle{empty}

\vspace{-0.8cm}

\begin{center}
\scriptsize
Institute for Nuclear Research and Nuclear Energy, \\
Tsarigradsko Chaussee 72, BG-1784 Sofia, Bulgaria \\
mitov@inrne.bas.bg
\end{center}

I would like to explain first how I came
to this subject.
About four years ago I became interested
in the possibility of
extending
the
algebraic methods of conformal
field theory to more general quantum field
theories and especially to perturbative
quantum field theory.
It is known that
the
models of conformal field
theory admit
a
purely algebraic description.
This happens not only in two space--time
dimensions, where we have an infinite dimensional
group of conformal transformations but
also in higher dimensions.
The possibility of a purely algebraic description of
conformally invariant quantum fields is
mainly due to the simple type of the
singularities of the products of these fields.
These singularities, at least in the presence
of the most strong conformal invariance that
is so called global conformal invariance\footnote{%
The notion of ``global conformal invariance'' was introduced
almost 10 years ago in the paper
``Rationality of conformally invariant local correlation functions on compactified Minkowski space''
by N.M. Nikolov and I.T. Todorov
(Commun. Math. Phys. {\bf 218} (2001) 417-436).},
are singularities on light--like distances
of rational type.
The (purely) algebraic structure, which describes
a Globally Conformal Invariant QFT is called
vertex algebra
but it could be also called Operator Product Expansion (OPE) Algebra
because it just describes the OPE of all local fields in the theory
in algebraic terms.

In such a way a question arises whether
we can find more wide algebra of functions,
which would be sufficient for describing the
singularities of the products of quantum fields
in a more general situation.
We may look for this algebra as a certain
differential--algebraic extension of the
algebra of rational functions with light--cone
singularities.
We have somewhat analogous situation
to the situation in
abstract algebra, where
one starts with the field of rational numbers and then
introduces the algebraic numbers and even further,
the ``differential algebraic numbers'' known as periods.

The problem of finding of appropriate function spaces
for describing
the correlation functions in quantum field theory I have investigated
first in the direction of developing general
theory of OPE algebras, their deformation
theory and its relation to perturbative quantum
field theory.
Now, I am preparing the results of this my
long research, which are still not complete.
But at some point I decided to look at
the problem from the point of view of
perturbative quantum field theory and especially
from the point of view of the Gell--Mann--Low renormalization group
and then I found much easier and even
nice picture.
In this talk I will present this my research.

So, this talk will be devoted to
renormalization theory in configuration
spaces for Euclidean perturbative quantum
field theory.
Such an approach, named after H.~Epstein and V.~Glaser,
was long ago developed
on
Minkowski space and even on pseudo--Riemann manifolds
but it was not systematically
considered on
Euclidean space.
I will make some introductory remarks
to this subject:
why renormalization in configuration spaces?

The renormalization theory is usually considered
in momentum space, where there are general
methods
to deal
with Feynman integrals.
On the other hand, the renormalization on
configuration space has a direct geometric
interpretation allowing generalization even
on manifolds.
This geometric
interpretation is completely lost in momentum
space, or at least it becomes rather implicit.
Another feature of the Epstein-Glaser
renormalization is that it is done for the
products of fields.
This also facilitates the generalization of
to
perturbation theory on manifolds
but still it has the disadvantage of being
rather complicated technically,
especially for concrete calculations.

One of the results I will present
is an analog of the Epstein--Glaser
approach, which is entirely stated in terms of
renormalization of integrals of functions.
It is an old idea that
renormalization is a problem
of extending distributions.
In this work we axiomatize these extension maps
and call them renormalization maps.
This approach then has the additional advantage
of being independent of concrete models
of quantum fields like
the
$\varphi^4$--theory or
quantum electrodynamics etc.
This is because we
just
consider the renormalization maps
as acting on certain function spaces
regardless of any model of perturbative QFT.
In this way we get rid of the
technical difficulties present in one or
other models, in other words, we separate them
from the renormalization problem.
Furthermore, reformulating the renormalization problem
as operations on spaces of functions,
makes
the
geometric interpretation possible.
This is because the function spaces on which we
define these renormalization maps are
the spaces of regular functions on certain domains.
Hence, the geometric properties of these domains
determine the ambiguity of the renormalization.
Our main goal in this work is to use this geometric characterization
of the renormalization ambiguity in order to derive
an algebraic algorithm for determining the Gell--Mann--Low renormalization group action,
i.e., the action of the one parameter group $\R^+$ on the space of coupling constants,
which is induced by the scaling transformations (in terms of formal diffeomorphisms).
In particular, we are interested in algebraic algorithms for calculating the perturbative expansions
of $\beta$--functions and anomalous dimensions.

\bigskip

So, the renormalization in the
approach I am presenting is introduced
by a system of linear maps each of them being applied on
a certain algebra of smooth functions
that are not globally defined
and as a result producing
everywhere defined distributions:
$$
\left\{
\begin{array}{c}
\text{algebra {$\PA_n$} of} \\
\text{non globally defined} \\
\text{smooth functions}
\end{array}
\right\}
\hspace{10pt}
\raisebox{-3pt}{$\mathop{\text{\Huge $\to$}}\limits^{\raisebox{3pt}{\large $\RMA_n$}}$}
\hspace{10pt}
\left\{
\begin{array}{c}
\text{space of} \\
\text{globally defined} \\
\text{distributions}
\end{array}
\right\} \,.
$$
I begin with the definition of the
algebras of regular functions on which we apply the renormalization maps.
Shortly speaking,
this is a sequence of algebras $\PA_2$, $\PA_3$, $\dots$, $\PA_n$, $\dots$
etc., where $\PA_n$ is an algebra of translation
invariant functions of $n$ vector arguments
and these functions one can think of as
coming from Feynman diagrams.
More precisely we assume that the
algebra $\PA_n$ is linearly spanned by finite
linear combinations of products of the form
$$
G \, = \,
\mathop{\prod}
\limits_{1 \, \leqslant \, j \, < \, k \, \leqslant \, n} \,
G_{jk} \bigl(\x_j-\x_k\bigr)
, \quad
\x_k \, \in \, \HES \, \equiv \, \R^D\,,
$$
where $\x_k$ $=$ $(x^1,\dots,x^D)$ are Euclidean vectors
and the functions $G_{jk}(\x)$ belong to the
algebra $\PA_2$, which is a subalgebra of
the algebra $\CI \bigl(\HES \mz\bigr)$ of the smooth functions
outside the origin.
One can think of the algebra $\PA_2$ as an
algebra containing the propagators of the theory.
In this way the algebras $\PA_n$ for $n>2$ are
entirely determined by the algebra $\PA_2$.

One technical assumption for the algebra $\PA_2$,
and thus for all other $\PA_n$, is that it is closed
with respect to multiplication of its elements by
polynomials as well as with respect to applying derivatives.

A main example is obtained by setting
$\PA_2$ to be the algebra of rational functions
of the form polynomials over powers of
the Euclidean square $\x^2$ $=$
$(x^1)^2$ $+$ $\cdots$ $+$ $(x^D)^2$,
$$
\PA_2 \, = \, \Bigl\{\frac{p(\x)}{(\x^2)^N} :
p(\x) \text{ -- polynomial}, N \, \in \, \N
\Bigr\}
\qquad (\x \, = \, \x_1-\x_2) \,.
$$
In this case $\PA_n$ is the algebra of
rational functions that are ratios of
translation invariant polynomials of $n$
vectors and powers of the product of
all Euclidean squares of the mutual
differences:
$$
\PA_n \, = \, \Biggl\{\frac{p(\x_1-\x_n,\dots,\x_{n-1}-\x_n)}{
\Bigr(\mathop{\prod}\limits_{j \, < \, k}(\x_j-\x_k)^2\Bigl)^N} :
p \text{ -- polynomial}, N \, \in \, \N
\Biggr\} \,.
$$
This example would correspond to the case when
we perturb free massless fields and consider
the mass terms in the Lagrangian as perturbations.
This is convenient if we wish to work algebraically
as much as possible.

Another remark: the algebra $\PA_n$ consists of
translation invariant functions,
which are regular on the so called configuration
space
$$
F_n \, = \, \bigl\{(\x_1,\dots,\x_n) \in \HES^n :
\x_j \, \neq \, \x_k \ \ (\forall j \neq k)\bigr\} \,.
$$
From the point of view of the algebraic
geometry $\PA_n$ is the ring of regular
functions on the quasiaffine manifold
that is the complement of union of quadrics
$$
F_{n;\hspace{1pt} \C} \, = \, \bigl\{(\x_1,\dots,\x_n) \in \C^{Dn} :
(\x_j - \x_k)^2 \neq 0 \ \ (\forall j \neq k)\bigr\} \,.
$$

Now, the renormalization maps we shall
define as linear maps
$$
\RMA_n \, : \, \PA_n \, \to \,
\DP \bigl(\HES^{\times n} \bigl/ \HES\bigr)
$$
without any requirement of continuity
with respect of some nontrivial topology.
These linear map are supposed to fulfill
the following axiomatic conditions $(r1)$--$(r4)$.

The first condition ($r1$) is the permutation symmetry:
$\RMA_n \bigl(\sigma^* G\bigr)$
$=$ $\sigma^* \RMA_n (G)$,
$\forall \sigma \in \Ss_n$,
where
$\sigma^* F (\x_1,\dots,\x_n)$ $:=$ $F (\x_{\sigma_1},\dots,\x_{\sigma_n})$.

I shall use the convention that for
every non-empty finite subset $S$ of
the set of natural numbers $\N$
we define an algebra
$\PA_S$ isomorphic to $\PA_n$ if
$n$ is the number of elements of $S$,
but $\PA_S$ is spanned by products of 2--point
functions of differences of vectors indexed
by the elements of the set $S$
(i.e.,
\(\prod_{j,\, k \, \in \, S,\ j \, < \, k} \,
G_{jk} \bigl(\x_j-\x_k\bigr)\))
Then we define also renormalization
maps
$$
\RMA_S \, : \, \PA_S \, \to \, \DP \bigl(\HES^S \bigl/ \HES\bigr)
$$
setting
$$
\RMA_S (G) \, = \, (\sigma^*)^{-1} \RMA_n \bigl(\sigma^* G\bigr)\,,
$$
where $\sigma : \{1,\dots,n\} \cong S$ and
$\sigma^* G (\x_1,\dots,\x_n)$ $:=$ $G (\x_{\sigma_1},\dots,\x_{\sigma_n})$.
I will often use these set--subscript
notations because of their convenience
for the related combinatorics
and everywhere further $S$ and $S'$ will
denote finite subsets of the set of
positive integers.

The next condition ($r2$) is preservation
of certain filtrations defined by the
notion of the scaling degree:
$$
\text{Scaling degree of } \RMA_n G \, \leqslant \,
\text{Scaling degree of } G \,.
$$
The scaling degree gives the rate of
the singularity for coinciding arguments.
In particular, we require that all the elements
of our algebras have finite scaling degrees.

The next condition ($r3$) requires
commutativity between the renormalization
maps and the multiplication by
polynomials
$$
\RMA_n \bigl(p G\bigr) \, = \, p \RMA_n G
, \quad
p = p(\x_1-\x_n, \dots, \x_{n-1}-\x_n) \text{ is a polynomial.}
$$
I consider in this property only polynomials
since I wish to work algebraically but
if we work on manifolds
then it is natural to require commutativity
between the renormalization maps and multiplication
by everywhere smooth functions.
In the latter case the above property becomes very natural
from geometric point of view since it allows us to make localization
(i.e., to use localization techniques like partition of unity).
We shall also see in what follows that the above property $(r3)$
is crucial for the reduction of our cohomological analysis
to de Rham cohomologies of configuration spaces.

The last requirement $(r4)$ is nonlinear and
relates recursively $\RMA_n$ with the renormalization
maps of lower order.
To introduce it I shall introduce first some
notations.
Let $\PRT$ be a partition of the set $S$ $=$ $\{j_1,\dots,j_n\}$
$$
\Bigl(\mathop{\bullet}\limits_{j_1} \mathop{\bullet}\limits_{j_2}
\mathop{\bullet}\limits_{j_3}\Bigr)
\cdots \Bigl( \cdots \mathop{\bullet}\limits_{j_k} \cdots \Bigr) \cdots
\Bigl(\cdots
\mathop{\bullet}\limits_{j_n}\Bigr) \,.
$$
Then we shall identify $\PRT$ also with the
equivalence relation on $S$ whose equivalence
classes coincide with the elements of $\PRT$.
This equivalence relation I shall denote by $\sim_\PRT$.
Then I define the following open subsets
of the Cartesian power $\HES^S$,
one for every $S$--partition $\PRT$:
$$
F_{\PRT} \, = \, \bigl\{(\x_{j_1},\dots,\x_{j_n}) \in \HES^S :
\x_j \, \neq \, \x_k \ \ (\forall j \nsim_{\PRT} k)\bigr\} \,.
$$
Also for a completely multiplicative elements
$$
G \, = \,
\mathop{\prod}
\limits_{\mathop{}\limits^{j,\,k \, \in \, S}_{j \, < \, k}} \,
G_{jk} \bigl(\x_j-\x_k\bigr)
$$
I introduce the decomposition,
$$
G_S \, = \,
G_{\PRT} \, \cdot \,
\mathop{\prod}\limits_{S' \, \in \, \PRT} G_{S'} \,,
$$
where
$G_{\PRT}$ contains all two--point functions
between nonequivalent points
$$
G_{\PRT} \, = \,
\mathop{\prod}
\limits_{\mathop{}\limits^{j \, \nsim_{\PRT} \, k}_{j \, < \, k}} \,
G_{jk} \bigl(\x_j-\x_k\bigr)
$$
and the remaining two-point functions are combined
into products
$$
G_{S'} \, = \,
\mathop{\prod}
\limits_{\mathop{}\limits^{j,\,k \, \in \, S'}_{j \, < \, k}} \,
G_{jk} \bigl(\x_j-\x_k\bigr) \,.
$$
Under these conventions the forth requirement
on the renormalization maps states that:
$$
\RMA_S G_S \vrestr{12pt}{F_{\PRT}} \, = \,
G_{\PRT} \, \cdot \,
\mathop{\prod}\limits_{S' \, \in \, \PRT} \RMA_{S'} G_{S'} \,.
$$
Here $G_{\PRT}$ is a multiplicator on $F_\PRT$.

From an axiomatic point of view it is
enough to impose the condition $(r4)$ only
for partitions with two elements.
By convention for elements $S'$ of the partition $\PRT$,
which contain one element we set $G_{S'}$ $=$ $1$.
Similarly, if the partition $\PRT$ consists of
one element we set $G_{\PRT}$ $=$ $1$.
Then as a particular case of ($r4$) we obtain that
$$
\RMA_n G \vrestr{12pt}{F_n} \, = \, G \,,
$$
i.e., the renormalization maps provides extensions
of smooth functions on the configuration spaces $F_n$
to everywhere defined distributions.

Let\hspace{-0.11pt} me\hspace{-0.11pt}
summarize\hspace{-0.11pt} the\hspace{-0.11pt} axiomatic\hspace{-0.11pt}
conditions\hspace{-0.11pt} on\hspace{-0.11pt} the\hspace{-0.11pt}
renormalization\hspace{-0.11pt} maps:

\medskip

($r1$) is the permutation symmetry;

($r2$) is the preservation of the filtrations;

($r3$) is commutativity of renormalization
maps with the multiplication by polynomials;

($r4$) is the recursive relation
$$
\RMA_S G_S \vrestr{12pt}{F_{\PRT}} \, = \,
G_{\PRT} \, \cdot \,
\mathop{\prod}\limits_{S' \, \in \, \PRT} \RMA_{S'} G_{S'} \,.
$$

\medskip

Now, I shall show briefly
how these
maps can be combined within the Euclidean
perturbative quantum field theory.\footnote{%
the material in this paragraph is not published in the paper
arXiv:0903.0187}
There we need to define products of
interactions
$$
I_1 (\x_1) \cdots I_n(\x_n),
$$
as quadratic forms on the Euclidean Fock space.
Here, every $I_k(\x)$
is a Wick polynomial of the basic fields $\varphi(\x)$
and its derivatives $\di^{\rR} \varphi (\x)$:
$$
I_k(\x) \, = \,
:\!
\text{Polynomial}
\bigl(\varphi (\x), \di_{\x} \varphi (\x), \dots\bigr) \!: \,.
$$
Note that $I_k (\x)$ is well defined only as a quadratic form
on the Euclidean Fock space
\textit{but it is not representable by operator
if its degree as a polynomial is larger than 1.}
The latter is in contrast to the situation on the Minkowski space,
where we have well defined Wick monomials due to the possibility
of multiplying Wightman distributions.
Thus, in order to define the product
$I_1 (\x_1)$ $\cdots$ $I_n(\x_n)$ of Wick monomials of Euclidean fields
we need renormalization.
To this end we formally decompose
$I_1 (\x_1)$ $\cdots$ $I_n(\x_n)$ by the Wick theorem:
$$
I_1 (\x_1) \cdots I_n(\x_n)
\, = \,
\mathop{\sum}\limits_{A_1,\dots,A_n} \, G_{A_1,\dots,A_n} (\x_1,\dots,\x_n)
:\!\Phi_{A_1} (\x_1) \cdots \Phi_{A_n} (\x_n)\!:\,,
$$
where the normal products are quadratic forms
smoothly dependent on $\x_1$, $\dots$, $\x_n$ and
$G_{A_1,\dots,A_n}$ are functions belonging to the algebra $\PA_n$.
The functions $G_{A_1,\dots,A_n}$ come from the Wick pairings
and they belong to the algebra $\PA_n$.
Then the renormalized product of interactions is defined as
\beqa
\bigl(I_1 (\x_1) \cdots I_n(\x_n)\bigr)^{\text{ren}}
\, = \,
\mathop{\sum}\limits_{A_1,\dots,A_n} \podr
\RMA_n \bigl(G_{A_1,\dots,A_n} \bigr) (\x_1,\dots,\x_n)
\nn \podr \times \,
:\!\Phi_{A_1} (\x_1) \cdots \Phi_{A_n} (\x_n)\!:\,.
\nonumber
\eeqa
In fact, we have a formula
\beqa
I_1 (\x_1) \cdots I_n(\x_n)
\, = \,
\mathop{\prod}\limits_{1 \, \leqslant \, j \, < \, k \, \leqslant \, n}
\podr
\exp \Biggl(
\mathop{\sum}\limits_{\rR,\sS} C_{\rR,\sS} \bigl(\x_j-\x_k\bigr)
\frac{\di}{\di \varphi_{\rR} (\x_j)} \frac{\di}{\di \varphi_{\sS} (\x_k)}
\Biggr)
\nn \podr \times \,
:\!I_1 (\x_1) \cdots I_n(\x_n)\!: \,.
\nonumber
\eeqa
$$
\varphi_{\rR} (\x) \, := \,
\di_{\x_1}^{\rR} \varphi (\x)
, \quad
C_{\rR,\sS} \bigl(\x_1-\x_2\bigr) \, := \,
\di_{\x_1}^{\rR} \di_{\x_2}^{\sS}
\La \varphi(\x_1) \varphi(\x_2) \Ra \,.
$$
This formula is convenient since the prefactor has the
multiplicative form used in the renormalization
recursion.

Let us come back to the axiomatic requirements
for the renormalization maps:

($r1$) permutation symmetry;

($r2$) preservation of the filtrations;

($r3$) commutativity of renormalization
maps with the multiplication by polynomials;

($r4$) recursive relation
$$
\RMA_S G_S \vrestr{12pt}{F_{\PRT}} \, = \,
G_{\PRT} \, \cdot \,
\mathop{\prod}\limits_{S' \, \in \, \PRT} \RMA_{S'} G_{S'} \,.
$$

In fact, I shall not consider these conditions as
a definition of renormalization maps but I shall
give a construction for these maps by means
of simpler objects and these I shall
consider as the exact definition of
renormalization.

To this end let me point out that the
open sets $F_{\PRT}$ that I defined in the
condition $(r4)$ form an open covering of
the complement of the total diagonal $\Delta_S$:
$$
\HES^S \backslash \Delta_S \, = \, \mathop{\bigcup}
\limits_{\mathop{}\limits^{\PRT \text{ is a proper}}_{\text{$S$--partition}}}
F_{\PRT} \,.
$$
Then we obtain a linear map
$$
\SRMA_S : \PA_S \to \DP_{temp}
\Bigl(\bigl(\HES^S \bigl/ \HES\bigr) \mz\Bigr)\,,
$$
where
the subscript ``temp'' means that we consider
distributions with a finite scaling degree.
(Note that
$\bigl(\HES^S \bigl/ \HES \bigr) \mz$ $\cong$
$\bigl(\HES^S \backslash \Delta_S \bigr) \bigl/ \HES$
and
a distribution on $\bigl(\HES^S \bigl/ \HES\bigr) \mz$
is the same as a translation invariant distribution on $\HES^S$
defined outside the total diagonal $\Delta_S$.)
I shall call these linear maps $\SRMA_S$
secondary renormalization maps and they
are completely determined by $\RMA_{S'}$ for
sets $S'$ containing less than $n$ elements.
Then to construct $\RMA_S$ we should compose $\SRMA_S$
with a linear map
$$
\PRMA_S : \DP_{temp} \Bigl(\bigl(\HES^S \bigl/ \HES\bigr) \mz\Bigr)
\to \DP \bigl(\HES^S \bigl/ \HES\bigr)\,,
$$
$$
\RMA_S \, = \, \PRMA_S \circ \SRMA_S \,,
$$
$$
\PA_S
\, \mathop{\longrightarrow}\limits^{\SRMA_S} \,
\DP_{temp} \Bigl(\bigl(\HES^S \bigl/ \HES\bigr) \mz\Bigr)
\, \mathop{\longrightarrow}\limits^{\PRMA_S} \,
\DP \bigl(\HES^S \bigl/ \HES\bigr) \,.
$$
These maps I shall call primary
renormalization maps.

The axiomatic conditions on $\PRMA_S$ are the
following.

\medskip

($p1$)
$\PRMA_S u \vrestr{12pt}{\bigl(\HES^S \bigl/ \HES\bigr) \mz}$ $=$ $u$,\
i.e. $\PRMA_S$ makes extension of distributions.

($p2$) Preservation of the filtrations.

($p3$) Orthogonal invariance.
(This is with respect to all Euclidean transformations of
$\HES^S \bigl/ \HES$.
It then will imply both, the permutation
symmetry of $\RMA_n$ and their orthogonal invariance with respect to $\HES$.)

($p4$) Commutativity with the multiplication by polynomials.

($p5$) If $u (\x) \in \DP \bigl(\HES^{S\backslash S'}\bigr)$
is a distribution supported at zero and $v(\y)$ is a distribution
belonging to $\DP_{temp} \Bigl(\bigl(\HES^{S'} \bigl/ \HES\bigr) \mz\Bigr)$
then
$$
\PRMA_S (u \otimes v) \, = \, u \otimes \PRMA_{S'} v \,.
$$

I shall not consider here the construction of
these renormalization maps but it can be
found in my paper (Sect.~2.5).

Before I proceed to the announced topic on
anomalies and cohomologies I will consider
briefly the problem of changing the renormalization.
The following statement holds:

\medskip

\noindent
{\bf Theorem.}\
{\it
Let $\{\PRMA_n\}_{n \, = \, 2}^{\infty}$
and $\{\PRMA_n'\}_{n \, = \, 2}^{\infty}$
be two systems of primary renormalization maps,
which define the systems
$\{\RMA_n\}_{n \, = \, 2}^{\infty}$
and $\{\RMA_n'\}_{n \, = \, 2}^{\infty}$
of renormalization maps, respectively.
Then for every finite $S \subset \N$ and a multiplicative element
$G_S \in \PA_S$ the following formula holds:
}
$$
\RMA_S' \, G_S \, = \,
\mathop{\sum}
\limits_{\mathop{}\limits^{\PRT \text{ is a}}_{\text{$S$--partition}}}
\Bigl(\RMA_{S/\PRT} \otimes \id_{\DP_{\PRT,0}}\Bigr)
\circ \NORM_{\PRT}
\left(\raisebox{12pt}{\hspace{-2pt}}\right.
G_{\PRT} \,
\mathop{\prod}\limits_{S' \, \in \, \PRT}
\Rdf_{S'} \, G_{S'}
\left.\raisebox{12pt}{\hspace{-2pt}}\right)\,.
$$
Here
\ \(
\Rdf_S \, = \, \bigl(\PRMA_S' - \PRMA_S\bigr) \circ
\SRMA_S{\hspace{-5pt}}' \hspace{2pt}\)\
are linear maps
$\PA_S \to \DP_{S,0}$ $:=$ $\DP \bigl[\Mbf{0} \in \HES^S \bigl/ \HES\bigr]$\
(distributions supported at the origin)
if $|S| > 1$,
otherwise,\ $\Rdf_{S'} \, G_{S'}$ $=$ $1$;
$S/\PRT$ $:=$ $\{\max \, S' : S' \in \PRT\}$;
$\NORM_{\PRT}$ is the linear map
\(
\NORM_{\PRT} :
\PA_{\PRT} \otimes \DP_{\PRT,0} \to \PA_{S/\PRT} \otimes \DP_{\PRT,0} \,.
\)

The details on this theorem can be found
in my paper (Sect.~2.6) and here I would like to point out
that the change of the renormalization is
characterized by a sequence of linear maps
$\Rdf_n : \PA_n \to \DP_{n,0}$, $\Rdf_1 = 1$,
satisfying the properties

\medskip

($c1$) permutation symmetry;

($c2$) preservation of the filtrations;

($c3$) commutativity with multiplications by polynomials.

\medskip

\noindent
The set of all such systems of linear maps
form a group with a multiplication
$$
\Rdf''_S \, G_S \, = \,
\mathop{\sum}
\limits_{\mathop{}\limits^{\PRT \text{ is a}}_{\text{$S$--partition}}}
\Bigl(\Rdf'_{S/\PRT} \otimes \id_{\DP_{\PRT,0}}\Bigr)
\circ \NORM_{\PRT}
\left(\raisebox{12pt}{\hspace{-2pt}}\right.
G_{\PRT} \,
\mathop{\prod}\limits_{S' \, \in \, \PRT}
\Rdf_{S'} \, G_{S'}
\left.\raisebox{12pt}{\hspace{-2pt}}\right).
$$
This group I called ``universal renormalization group''
and it corresponds to Stueckelberg renormalization group in our formalism.

When we apply the renormalization maps to a
perturbative quantum field theory with some fixed initial
field content we obtain a representation of
the above universal renormalization group in the
group of all formal diffeomorphisms
of the coupling constants that parameterize
all possible interactions for the fixed set
of initial fields.
A key role in the derivation of this representation
play the formulas
$$
\RMA_S' \, G_S \, = \,
\mathop{\sum}
\limits_{\mathop{}\limits^{\PRT \text{ is a}}_{\text{$S$--partition}}}
\Bigl(\RMA_{S/\PRT} \otimes \id_{\DP_{\PRT,0}}\Bigr)
\circ \NORM_{\PRT}
\left(\raisebox{12pt}{\hspace{-2pt}}\right.
G_{\PRT} \,
\mathop{\prod}\limits_{S' \, \in \, \PRT}
\Rdf_{S'} \, G_{S'}
\left.\raisebox{12pt}{\hspace{-2pt}}\right)
$$
and
\beqa
I_1 (\x_1) \cdots I_n(\x_n)
\, = \,
\mathop{\prod}\limits_{1 \, \leqslant \, j \, < \, k \, \leqslant \, n}
\podr
\exp \Biggl(
\mathop{\sum}\limits_{\rR,\sS} C_{\rR,\sS} \bigl(\x_j-\x_k\bigr)
\frac{\di}{\di \varphi_{\rR} (\x_j)} \frac{\di}{\di \varphi_{\sS} (\x_k)}
\Biggr)
\nn \podr \times \,
:\!I_1 (\x_1) \cdots I_n(\x_n)\!: \,,
\nonumber
\eeqa
which I have already presented.
It is important here the multiplicative
form of the prefactor and by a preliminary
investigations of mine, for the description of
the action of the universal renormalization group
by formal diffeomorphisms it would be
convenient to introduce a certain coalgebraic
structure in the space of all interactions.

I proceed to the anomalies in perturbative quantum field theory.
We speak about anomalies when a symmetry
of the unrenormalized, or bare, Feynman integrals
is broken after the renormalization.
It is clear that for this a main role plays
the absence of commutativity between the
renormalization maps and the action of
linear partial differential operators.

By construction the renormalization maps
commute with the multiplication by polynomials
and then what remains as a source for the anomalies
are the commutators
$$
\bigl[\di_{x^{\mu}_k},\RMA_n\bigr] \, = \,
\bigl[\di_{x^{\xi}},\RMA_n\bigr]
, \quad
\xi \, = \, (k,\mu)\,.
$$
Let us denote
$\Lccl_{n;\, \xi}$ $:=$ $\bigl[\di_{x^{\xi}},\RMA_n\bigr]$
and apply the main formula
$$
\RMA_n \, = \, \PRMA_n \circ \SRMA_n \,.
$$
We obtain a decomposition,
\beqa
\Lccl_{n;\, \xi} \, = \podr
\Sccl_{n;\, \xi} \, + \, \SLccl_{n;\, \xi}
\qquad (n>2), \qquad \Lccl_{2;\, \xi} \, \equiv \, \Sccl_{2;\, \xi}
\, , \quad
\nonumber \\
\Sccl_{n;\, \xi} \, := \podr
\bigl[\di_{x^{\xi}},\PRMA_n\bigr] \circ \SRMA_n
\qquad (n>2)
\, , \quad
\nonumber \\
\SLccl_{n;\, \xi} \, := \podr
\PRMA_n \circ \bigl[\di_{x^{\xi}},\SRMA_n\bigr]
\qquad (n>2)
\, , \quad
\nonumber
\eeqa
where the important point is that $\SLccl_{n;\, \xi}$ are recursively
determined due to the commutator with
the secondary renormalization maps:
it will
produce at least one delta function or
its derivatives supported on some partial diagonal
and after that we have a renormalization on
less than $n$ variables.

On the other hand, $\Sccl_{n;\, \xi}$ are linear maps
$\PA_n \to \DP_{n,0}$ $=$ $\DP \bigl[\Mbf{0} \in \HES^n \bigl/ \HES\bigr]$,
i.e. they produce distributions supported at zero,
or, in other words, total delta functions and derivatives.
Thus, $\Sccl_{n;\, \xi}$ are simpler linear maps
than $\Lccl_{n;\, \xi}$.
Our idea is to characterize $\Sccl_{n;\, \xi}$ by
cohomological equations such that the
ambiguity in their solutions to be exactly
corresponding to the renormalization ambiguity.
Such a system of equations is the following:
\beqa
\podr \!\!
\bigl[\di_{x^{\xi}},\Sccl_{2;\, \eta}\bigr]
-
\bigl[\di_{x^{\eta}},\Sccl_{2;\, \xi}\bigr] \, = \, 0 \,,
\nn
\podr \!\!
\bigl[\di_{x^{\xi}},\Sccl_{n;\, \eta}\bigr]
-
\bigl[\di_{x^{\eta}},\Sccl_{n;\, \xi}\bigr]
\nn \podr \!\! \hspace{20pt}
= \,
- \, \bigl[\di_{x^{\xi}},\PRMA_n\bigr] \circ \bigl[\di_{x^{\eta}},\SRMA_n\bigr]
\, + \,
\bigl[\di_{x^{\eta}},\PRMA_n\bigr] \circ \bigl[\di_{x^{\xi}},\SRMA_n\bigr]
\, \quad (n > 2) \,.
\nonumber
\eeqa
It can be verified by straightforward computations.
We see that the right hand side looks like
a differential of a one--form.
Later we shall reduce this expression to
such a differential.
For the right hand side it is important to note that
it is determined by the renormalization recursion.
The reason is the same as above:
the second commutators in both
terms contains a secondary renormalization map
and then they produce at least one delta function
or its derivatives supported at some partial
diagonal;
this reduces the number
of points and thus also the order of the
renormalization.
The main result here is

\medskip

\noindent
{\bf Theorem.}\
{\it
Let $n>2$; given a system of primary renormalization maps
$\PRMA_2,$ $\PRMA_3,$ $\dots,$ $\PRMA_n$
(which therefore determine renormalization maps
$\RMA_2,$ $\RMA_3,$ $\dots,$ $\RMA_n$),
let $\{\Sccl_{n;\, \xi}\}_{\xi}$ be defined by
$\PRMA_2,$ $\PRMA_3,$ $\dots,$ $\PRMA_n$ and
$\{\Sccl_{n;\, \xi}'\}_{\xi}$ be a solution of the cohomological equations,
which differs from $\{\Sccl_{n;\, \xi}\}_{\xi}$ by an
\EMPHTH{exact solution}, i.e., the difference
$\Sccl_{n;\, \xi}'$ $-$ $\Sccl_{n;\, \xi}$ is of a form
$$
\Sccl'_{n;\, \xi} - \Sccl_{n;\, \xi} \, = \,
\bigl[\di_{x^{\xi}},\RDF_n\bigr]
$$
($\xi=1,\dots,D(n-1)$), for some linear map
$$
\RDF_n :
\PA_n
\to
\DP_{n,0} \,.
$$
Then there exists a primary renormalization map $\PRMA_n'$,
which together with $\PRMA_2,$ $\dots,$ $\PRMA_{n-1}$ determines
a system of renormalization maps
$\RMA_2,$ $\dots,$ $\RMA_{n-1}$ and $\RMA_n'$
and a primary renormalization cocycle coinciding with
$\{\Sccl_{n;\, \xi}'\}_{\xi}$.
}

\medskip

In this way the set of all renormalizations
determines a cohomology class and the above
equations fix this cohomology class.
Another important corollary is that if
the cohomologies related to the differential in
the right hand side are zero, then every
solution of these equations would correspond
to some renormalization map and there will be no need
to find what is exactly this renormalization map.
Unfortunately, if one starts with the algebras of
rational functions, which we have set in the beginning then these
cohomologies are nonzero, although they are finite
dimensional since they correspond to de Rham
cohomologies of the so called Bernstein $D$-modules.
So, this is exactly the reason why
we need to introduce transcendental extensions in order to describe
the anomalies in perturbative quantum field theory and in particular,
for calculating beta functions.

I shall now give a reduction of the objects,
which we have to determine by the cohomological
equations to simpler objects.
These were the maps $\Sccl_{n;\, \xi}$
and I will call them (primary) renormalization cocycles.
An important feature of these maps $\Sccl_{n;\, \xi}$ is
that they commute with the multiplication by
polynomials.
In particular,
$\bigl[x^{\eta}, \Sccl_{n;\, \xi}\bigr]$ $=$ $0$, i.e.,
$x^{\eta} \Sccl_{n;\, \xi} G$ $-$
$\Sccl_{n;\, \xi} (x^{\eta} G)$$=$ $0$.
It follows then that $\Sccl_{n;\, \xi}$ has the following form
$$
\Sccl_{n;\, \xi} (G) \, = \,
\mathop{\sum}\limits_{\rr \, \in \, \N_0^{N}}
\ \frac{(-1)^{|\rr|} \, }{\rr!} \
\sccl_{n;\, \xi} \bigl(\xx^{\rr} \hspace{1pt} G\bigr) \,
\delta^{(\rr)} (\xx)\,,
$$
where $\sccl_{n;\, \xi}$ are linear functionals on $\PA_n$
belonging to the subspace\footnote{%
We changed the notation $\PA_n^{\hspace{1pt}\bullet}$ used in the paper arXiv:0903.0187 to $\PA_n^{\du}$.}
$$
\PA_n^{\du}
\, := \,
\bigl\{\Gamma \in \PA\hspace{2pt}_{\hspace{-2pt}n}' : \exists N \text{ s.t. }
\Gamma \bigl(G\bigr) \, = \, 0 \text{ if } \Scdeg \, G \, \leqslant \, N
\bigr\}.
$$
Because of the condition
on the linear functionals to belong to the subspace $\PA_n^{\du}$
the sum in the above formula is finite.
Moreover, the correspondence
$\Sccl_{n;\, \xi}$ $\leftrightarrow$ $\sccl_{n;\, \xi}$ is
one--to--one and the commutator
$\bigl[\di_{x^{\eta}}, \Sccl_{n;\, \xi}\bigr]$
is transformed to the dual
derivative $-\sccl_{n;\, \xi} \circ \di_{x^{\eta}}$
of the linear functional $\sccl_{n;\, \xi}$.
In this way we reduce the description of every map $\Sccl_{n;\, \xi}$
to one linear functional $\sccl_{n;\, \xi}$ (defined on the same domain as $\Sccl_{n;\, \xi}$).
This is the point where the condition $(r3)$ plays a crucial role,
as I mentioned before.
Let us organize $\sccl_{n;\, \xi}$ into a $1$--form
$$
\bsccl_n \, = \,
\mathop{\sum}\limits_{\xi \, = \, 1}^N
\sccl_{n;\, \xi} \, dx^{\xi}
\, \in \, \OM{1} \bigl(\PA_n^{\du}\bigr)
\, := \,
\PA_n^{\du} \otimes \Lambda^1 \HES \,,
$$
with coefficients in the differential module
$\PA_n^{\du}$.
Then the cohomological equations can be written in
terms of these linear functionals as
\beqa
d \hspace{1pt} \bsccl_2 \, = \podr 0
\, , \qquad
\nn
d \hspace{1pt} \bsccl_n \, = \podr
\mathop{\sum}\limits_{m \, = \, 2}^{n-1} \,
\bsccl_{n-m+1}
\, \spwedge \,
\bsccl_m
\qquad (n > 2)
\,,
\nonumber
\eeqa
where $\spwedge$ is
a \textit{non--associative} and \textit{non--graded--commutative}
bilinear operation defined in the following way,
\beqa
\bigl(
\bsccl_{n-m+1}
\spwedge \,
\bsccl_m
\bigr)
\bigl(G_S\bigr)
\hspace{8pt}
\podr \nn
\, = \,
\mathop{\sum}
\limits_{\mathop{}\limits^{S' \, \subsetneq \, S}_{|S'| \, = \, m}}
\mathop{\sum}\limits_{\rr' \, \in \, \N_0^{N'}}
\podr
\frac{1}{\rr'!} \
\bsccl_{S/S'}
\Bigl(
\Mbf{\di}_{\xx{}'}^{\rr'} \hspace{1pt} G_{\PRT(S')}\vrestr{12pt}{\xx'=0}
\Bigr)
\wedge
\,
\bsccl_{S'}
\bigl(\xx{}'{}^{\rr'} \hspace{1pt} G_{S'}\bigr)
.
\nonumber
\eeqa
Still one can show that the latter operation is a graded \textit{pre--Lie} operation.
If we further organize the sequence $\{\bsccl_n\}$ into
a single object the cohomological equations read even
more compactly,
$$
d \underline{\bsccl} \, = \, \underline{\bsccl} \spwedge \underline{\bsccl}
\,,
$$
and this shows their integrability
since this is the form of Maurer-Cartan equations
in a graded differential Lie algebra.

Now, I shall consider the problem of
determining the cohomology spaces
$\HOM{1} \bigl(\PA_n^{\du}\bigr)$ of the complex
$\OM{k} \bigl(\PA_n^{\du}\bigr)$
$:=$ $\PA_n^{\du} \otimes \Lambda^k \HES$,
which control the ambiguity of the solutions of
the cohomological equations.
I have shown in my paper (Theorem 3.4 of Sect. 3.2) that we
have the following natural duality:
$$
\HOM{1} \bigl(\PA_n^{\du}\bigr)
\, \cong \,
\Bigl(\HOM{D(n-1)-1} \bigl(\PA_n\bigr)\Bigr)'\,,
$$
where $D(n-1)$ is the top degree.
Let me remind you that the algebra $\PA_n$
coincides with the algebra of regular functions
on the complement of the union of quadrics
$$
F_{n;\hspace{1pt} \C} \, = \, \bigl\{(\x_1,\dots,\x_n) \in \C^{Dn} :
(\x_j - \x_k)^2 \neq 0 \ \ (\forall j \neq k)\bigr\}
$$
modulo translations.
And so,
$$
\HOM{1} \bigl(\PA_n^{\du}\bigr)
\, \cong \,
\Bigl(\HOM{D(n-1)-1}_{DR} \bigl(F_{n;\hspace{1pt} \C} \bigl/ \C^D\bigr)\Bigr)'.
$$
If we work instead of with the algebras $\PA_n$
with (translation invariant) $\CI$--functions on
$$
F_n \, = \, \bigl\{(\x_1,\dots,\x_n) \in \HES^n :
\x_j \, \neq \, \x_k \ \ (\forall j \neq k)\bigr\} \,
$$
then we would of obtain
the de Rham cohomologies of the space
$F_n \bigl/ \HES$
of codegree $1$.
It is known that they are zero for $n\geqslant 3$
(see, e.g., {\it E.R. Fadell and S.Y. Husseini,
Geometry and Topology of Configuration Spaces, Springer}).
In this way we arrive to the idea to look for
an intermediate differential--algebraic extension,
$$
\PA_n \, \subseteq \, \PAE_n \, \subseteq \, \CI_{temp} \bigl(\HF_n\bigr),
$$
which would trivialize the  cohomologies
$$
\HOM{D(n-1)-1} \bigl(\PAE_n\bigr) \, = \, 0 \,.
$$

Let me mention a strategy for solving the cohomological equations:
\beqa
d \hspace{1pt} \bsccl_2 \, = \podr 0
\, , \qquad
\nn
d \hspace{1pt} \bsccl_n \, = \podr
\mathcal{F}_n \bigl[\bsccl_1,\dots,\bsccl_{n-1}\bigr]
\, = \,
\mathop{\sum}\limits_{m \, = \, 2}^{n-1} \,
\bsccl_{n-m+1}
\ \spwedge \
\bsccl_m
\qquad (n > 2)
\,,
\nonumber
\eeqa
for
\(
\bsccl_n \, = \,
\mathop{\sum}_{\xi}\
\sccl_{n;\, \xi} \, dx^{\xi}
\), where
$\sccl_{n;\, \xi} : \PA_n \to \R$,
which then, let me remind you, would determine the renormalization cocycles:
$$
\Sccl_{n;\, \xi} \, := \,
\bigl[\di_{x^{\xi}},\PRMA_n\bigr] \circ \SRMA_n \, = \,
\mathop{\sum}\limits_{\rr}
\ \frac{(-1)^{|\rr|} \, }{\rr!} \
\bigl(\sccl_{n;\, \xi} \circ \xx^{\rr}\bigr) \
\delta^{(\rr)} (\xx) \,.
$$
Assume:
$$
\exists \ \
K_n \, : \, \OM{D(n-1)-1} \bigl(\PAE_n\bigr) \to
\OM{D(n-1)-2} \bigl(\PAE_n\bigr)
\, , \quad
K_n \ \circ \ d + d \ \circ \ K_n \ = \ \id \,.
$$
Then a solution of the cohomological equations is:
$$
\bsccl_n \ = \
\mathcal{F}_n \ \circ \ K_n \,.
$$

Let me briefly demonstrate how all the above ideas look
in two space--time dimensions.
There the quadrics are reducible and
introducing the ``Euclidean light--cone coordinates'':
$$
(x^1,x^2) \, \leftrightarrow \, (z,w): \quad
z \, := \, x^1+i\, x^2 \quad \text{and} \quad  w \, := \, x^1-i\, x^2
$$
we obtain decompositions for the algebras $\PA_n$:
\beqa
\PA_n \, \cong \podr
\Q \bigl[z_1,\dots,z_{n-1}\bigr]
\Bigl[\Bigl(\mathop{\prod}\limits_j z_j\Bigr)^{-1}
\Bigl(\mathop{\prod}\limits_{j \, < \, k} (z_j-z_k)\Bigr)^{-1}\Bigr]
\nn \podr \otimes \
\Q \bigl[w_1,\dots,w_{n-1}\bigr]
\Bigl[\Bigl(\mathop{\prod}\limits_j w_j\Bigr)^{-1}
\Bigl(\mathop{\prod}\limits_{j \, < \, k} (w_j-w_k)\Bigr)^{-1}\Bigr] \,.
\nonumber
\eeqa

There is a result
({\it F.C.S. Brown,
Multiple zeta values and periods of moduli spaces $\mathfrak{M}_{0,n}$,
math/0606419})
stating that the algebra of the so called multiple polylogarithms
provides a differential extension
\beqa
\Q \bigl[z_1,\dots,z_{n-1}\bigr]
\Bigl[\Bigl(\mathop{\prod}\limits_j z_j\Bigr)^{-1}
\Bigl(\mathop{\prod}\limits_{j \, < \, k} (z_j-z_k)\Bigr)^{-1}\Bigr]
\nn
\subset \
\text{Multiple Polylogs} \bigl(z_1,\dots,z_{n-1}\bigr)\,,
\nonumber
\eeqa
which trivializes all the de Rham cohomologies.
Thus, we set:
\beqa
& \hspace{-14pt}
\PAE_n \, =
& \nn & \hspace{-14pt}
\Bigl(
\hspace{-1pt}
\text{Multiple\hspace{-1.5pt} Polylogs} \bigl(z_1,\dots,z_{n-1}\bigr)
\otimes
\text{Multiple\hspace{-1.5pt} Polylogs} \bigl(w_1,\dots,w_{n-1}\bigr)
\hspace{-1pt}
\Bigr)^{\mathop{}\limits^{\text{\tiny Monod--}}_{\text{\tiny romy}}}
\hspace{-3pt},
&
\nonumber
\eeqa
which in fact, requires an additional extension of the scalars:
$$
\Q \ \subset \ \text{Ring of multiple zeta values.}
$$
In this way
the linear functionals that determine the renormalization
cocycles would take values in the ring of multiple zeta values.
Since these maps are algebraically related
to the Gell--Mann--Low renormalization group action,
and in particular, to the series of the beta functions and anomalous dimensions,
it follows that the coefficients of the latter series
will be multiple zeta values
for any theory in two space--time dimensions.

Now the problem in higher dimensions looks like to find an
higher dimensional analog of the multiple polylogarithms:
$$
\Q \bigl[\x_1,\dots,\x_{n-1}\bigr]
\Bigl[\Bigl(\mathop{\prod}\limits_j \x_j^2\Bigr)^{-1}
\Bigl(\mathop{\prod}\limits_{j \, < \, k} (\x_j-\x_k)^2\Bigr)^{-1}\Bigr]
\,
\mathop{\subset}\limits^{\text{?}}
\,
\PAE_n \,.
$$
I have recently found a candidate for a such an extension
related to the problem of inverting the Laplace operator
on rational functions.
One needs such an extension for developing
a rigorous notion of perturbative operator product expansion algebras
and solving there field equations.
The extension $\PAE_n$ I obtained requires again only
the extension of the scalars from the field of rational numbers to
the ring of multiple zeta values.
So, it seems from this point of view that such a transcendental extension
is sufficient for the purposes of the perturbative quantum field theory.

\medskip

As a conclusion I would like to mention that
the renormalization in configuration spaces provides
a geometric insight to the problem what
are the transcendental extensions, which we need
for the function spaces that would be appropriate
for the description of the correlation functions in perturbative
quantum field theory.
It is important also to stress that
we play with the renormalization ambiguity in order to
find an algorithm for the Gell--Mann--Low renormalization group action,
on the space of all possible interactions.
Of course,
it would be trivial to play with the full renormalization ambiguity in order
to fix the series of the beta function in a particular theory.
With our method we intend to fix simultaneously the structure
of infinite number of formal power series including
not only the beta functions for all theories but also
the series of the anomalous dimensions.

Based on the additional results on
perturbative operator product expansion algebras,
which I have mentioned the above,
I would conjecture also that:
the coefficients of the beta functions in any perturbative
quantum field theory
on even space--time dimensions are multiple zeta values.

\medskip

\noindent
\textbf{Acknowledgments.}
I am grateful to Professor R. Stora for his critical remarks to this work.
I am also grateful to Professor I. Todorov for his comments.
I am grateful to the organizers of the Conference on Algebraic and Combinatorial Structures
in Quantum Field Theory in Carg\`ese and to
the Institut d'\'Etudes Scientifiques de Carg\`ese for the support and the hospitality.
This work was partially supported by Bulgarian NSF grant DO 02--257
and French--Bulgarian project Rila under the contract Egide -- Rila N112.

\end{document}